\documentclass[12pt]{iopart}
\usepackage[sorting=none, style = numeric-comp,url=false,doi=false,isbn=false,date=year ]{biblatex} 
\usepackage{graphicx}
\usepackage{xcolor}
\usepackage{lineno}
\usepackage{subcaption}
\usepackage{upgreek}
\usepackage{iopams}

\begin{document}

\title{Ultra-broadband non-degenerate guided-wave bi-photon source in the near and mid-infrared}

\author{F. Roeder,$^{1,2,*}$ A. Gnanavel,$^{1,2}$ R. Pollmann,$^{1,2}$ O. Brecht,$^{1,2}$ M. Stefszky,$^{1,2}$ L. Padberg,$^{1,2}$ C. Eigner,$^{2}$ C. Silberhorn,$^{1,2}$  B. Brecht$^{1,2}$}

\address{1 Paderborn University, Integrated Quantum Optics, Warburger Str. 100, 33098 Paderborn, Germany\\
2 Paderborn University, Institute for Photonic Quantum Systems (PhoQS), Warburger Str. 100, 33098 Paderborn, Germany}
\ead{franz.roeder@uni-paderborn.de}
\vspace{10pt}
\begin{indented}
\item[]\date{\today}
\end{indented}

\begin{abstract}
The latest applications in ultrafast quantum metrology require bright, broadband bi-photon sources with one of the photons in the mid-infrared and the other in the visible to near infrared. However, existing sources based on bulk crystals are limited in brightness due to the short interaction length and only allow for limited dispersion engineering. Here, we present an integrated PDC source based on a Ti:LiNbO$_3$ waveguide that generates broadband bi-photons with central wavelengths at $860\,\mathrm{nm}$ and $2800\,\mathrm{nm}$. Their spectral bandwidth exceeds $25\,\mathrm{THz}$ and is achieved by simultaneous matching of the group velocities and cancellation of group velocity dispersion for the signal and idler field. We provide an intuitive understanding of the process by studying our source's behaviour at different temperatures and pump wavelengths, which agrees well with simulations.   
\end{abstract}

\section{Introduction}

Broadband sources of non-degenerate bi-photons are required for quantum metrology applications, especially measurements with undetected photons. Here, such a source can be used as the active optical element within a so-called SU(1,1)-interferometer or nonlinear interferometer \cite{Yurke1986, Mukamel2020, Chekhova2016, Manceau2017}.
At the same time, sources with strong time-frequency entanglement and high spectral bandwidths provide short correlation times which are key for ultrafast quantum spectroscopy applications, such as quantum optical coherence tomography, Fourier transform infrared spectroscopy with undetected photons, or entangled two-photon absorption \cite{Ndagano2022, Yepiz-Graciano2020, katamadze2024, Kolenderska2020, Tashima_24,Lindner2021, Kaufmann2022,Landes2021, Raymer2022, Dayan2004, Shaked2014,Lemos2014, Toepfer2022, Riazi2019, Paterova2018, Kviatkovsky2020}.
Recent sources of non-degenerate broadband quantum light are based on bulk nonlinear materials that employ group-velocity matching \cite{Ramelow2019}, are poled aperiodically \cite{Fraine2012} or employ ultra-thin crystals, thus relaxing the phase-matching condition \cite{Chekhova2019}. However, these sources are limited in brightness and often do not provide collinear emission of the generated photons. The use of long periodically poled nonlinear waveguides provides a means to overcome these limitations \cite{Tanzilli2001}. Long waveguides, however, are usually associated with narrow spectra. To achieve ultra-broadband emission from waveguides, we must employ higher-order dispersion engineering techniques. As a first step in that direction we showed in previous works that group-velocity matching of signal and idler photons in a periodically poled Ti:LiNbO$_3$ waveguide can provide spectral bandwidths of more than $7\,\mathrm{THz}$ with correlation times below $100\,\mathrm{fs}$ and a spectral brightness exceeding $10^{6} \frac{\mathrm{pairs}}{\mathrm{s\cdot mW \cdot GHz}}$ \cite{Pollmann2024, Roeder2024}.\\
Here, we present a PDC source that features correlation times around $25\,\mathrm{fs}$, resulting from more than $25\,\mathrm{THz}$ of spectral bandwidth and that generates  highly non-degenerate bi-photons at central wavelengths of $860\,\mathrm{nm}$ and $2800\,\mathrm{nm}$. We achieve this via simultaneous matching of the signal and idler group velocities as well as operating at a point of zero total group velocity dispersion for these fields. Operating at this zero group velocity dispersion point is a technique often found in fiber-based sources that allow for flexible dispersion engineering, but exhibit lower nonlinear coefficients \cite{Smith2009, Chen2017, Zhu2013, Smirnov2024} due to the fact that these sources typically utilize the $\chi^{(3)}$ nonlinearity. Limitations in brightness can be overcome in integrated waveguides that employ the $\chi^{(2)}$ component and thereby inherently provide a higher brightness. First demonstrations in the novel platform of thin-film lithium niobate showed the potential to use dispersion engineering for achieving broadband single photon generation \cite{Williams2024, Fang24}. Alternatively, one can also use the long-established Ti:LiNbO$_3$ waveguide platform, which is more suitable for applications that couple to free-space or fibers, and where fabrication is both standardised and repeatable.

In this paper, we first present our method for engineering the PDC process and the necessary conditions for ultra-broadband bi-photon generation. We then introduce the setup for the characterization of the generated PDC signal photons and compare the measured spectra to our simulations. We investigate the dependence of the emission on the waveguide temperature as well as the pump wavelength. Thereafter, we evaluate the change in maximally achievable bandwidth from the source when operating at pump wavelengths away from the working point.

\section{PDC Process Engineering}

The spectral characteristics of a PDC waveguide source pumped by a continuous wave (cw) laser are given by the so-called joint spectral amplitude (JSA) \cite{Grice1997}:

\begin{equation}
	f(\Delta \omega) = \mathrm{sinc}\left(\frac{\Delta \beta (\Delta \omega)L}{2}\right)e^{i\frac{\Delta \beta(\Delta \omega)L}{2}}.
\end{equation}

Here, $\Delta \omega = \omega_s - \Omega_s = -(\omega_i-\Omega_i)$ is the signal (idler) frequency detuning from the central frequency $\Omega_s$ ($\Omega_i$). The sign change reflects the fact that signal and idler energies must add up to the pump energy such that $\hbar \Omega_p=\hbar \omega_s + \hbar \omega_i$, with $\Omega_p$ being the frequency of the pump laser. $L$ is the length of the waveguide. The exponential term in this expression contains the phase caused by the dispersion of the waveguide and $\Delta \beta$ is the phase mismatch between the pump and the generated signal and idler fields. This phase mismatch $\Delta \beta$ can be expanded as \cite{Uren2005}:

\begin{eqnarray}
    &\Delta \beta(\omega_s,\omega_i) = \beta_p(\omega_s+\omega_i) - \beta_s(\omega_s)-\beta_i(\omega_i)-\frac{2\pi}{\Lambda}\\ & \approx \Delta \beta^{(0)} + (\kappa_s - \kappa_i)\Delta \omega + \frac{1}{2} (\eta_s + \eta_i)\Delta \omega^2 - \eta_p \Delta \omega^2 + O(\Delta \omega^3)
\label{eq:expansion}
\end{eqnarray}

Here, the 0-th order phase mismatch $\Delta \beta^{(0)} = \beta_p(\Omega_s+\Omega_i)-\beta_s(\Omega_s)-\beta_i(\Omega_i)-\frac{2\pi}{\Lambda}$ is set to zero by an appropriate choice of the poling period $\Lambda$ to ensure phase matching for the central frequencies. The terms $\kappa_{s,i}=\left(\frac{\partial \beta_p}{\partial \omega}\bigg|_{\Omega_s + \Omega_i}- \frac{\partial \beta_{s,i}}{\partial \omega}\bigg|_{\Omega_{s,i}}\right)$ are related to the group velocities (GV) of the signal and idler photons, respectively, while the terms $\eta_{s,i} = \left(\frac{\partial^2\beta_p}{\partial \omega^2}\bigg|_{\Omega_s+\Omega_i} - \frac{\partial^2 \beta_{s,i}}{\partial \omega^2}\bigg|_{\Omega_{s,i}}\right)$ are related to their group velocity dispersion (GVD), in both cases relative to the pump field, for which $\eta_{p} = \frac{\partial^2\beta_p}{\partial \omega^2}\bigg|_{\Omega_p}$.
In the case of a cw pump, the contributions with derivatives containing $\beta_p$ cancel each other. Since the bandwidth of the generated bi-photon state is set by the width of the phase matching function, it is crucial to set the phase mismatch to zero in all these higher orders for a broad range of frequencies. To this end, we expand the phase mismatch up to second order in the frequency detuning $\Delta \omega$:

\begin{equation}
    \Delta \beta (\Delta\omega) \approx \left( - \frac{\partial \beta_s}{\partial \omega}\bigg|_{\Omega_s} + \frac{\partial \beta_i}{\partial \omega}\bigg|_{\Omega_i} \right) \Delta \omega - \frac{1}{2} \left( \frac{\partial^2 \beta_s}{\partial \omega^2}\bigg|_{\Omega_s} + \frac{\partial^2 \beta_i}{\partial \omega^2}\bigg|_{\Omega_i} \right) \Delta \omega^2
\end{equation}

The first order of the phase mismatch can be cancelled, if the signal and idler GVs are matched, i.e. $\left(-\frac{\partial \beta_s}{\partial \omega}\bigg|_{\Omega_s}+\frac{\partial \beta_i}{\partial \omega}\bigg|_{\Omega_i}=0\right)$. This condition is called GV matching which has already been exploited for dispersion engineering in our group \cite{Pollmann2024}. In addition, by ensuring that the GVD of signal and idler has equal magnitude but opposite sign, the second order of the phase mismatch can be cancelled, i.e. $\left(\frac{\partial^2 \beta_s}{\partial \omega^2}\bigg|_{\Omega_s}+\frac{\partial^2 \beta_i}{\partial \omega^2}\bigg|_{\Omega_i}=0\right)$. We refer to this as GVD cancellation. This condition is key to supporting the extreme bandwidths that are required for future applications in quantum metrology. 

\begin{figure}[ht!]
    \centering
    \includegraphics[width=\linewidth]{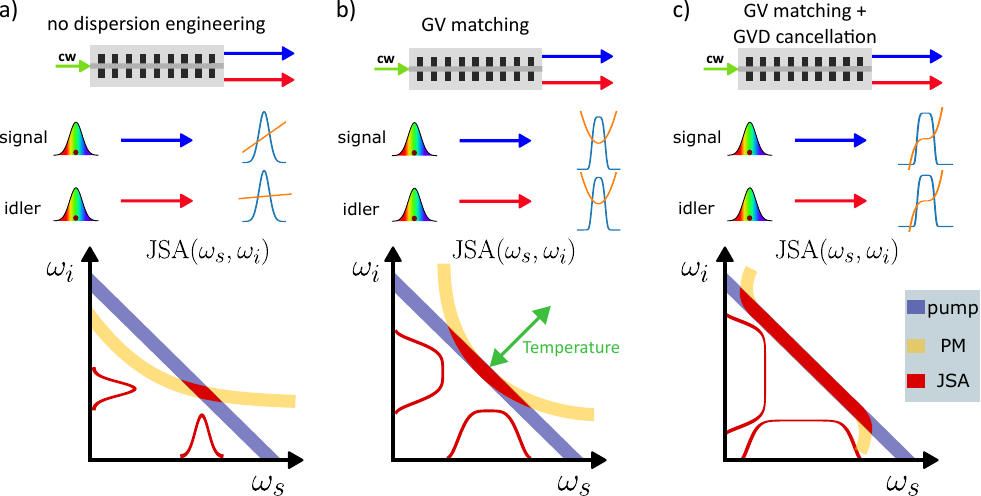}
    \caption{Dispersion engineering in a nonlinear waveguide: The dispersive effects on the generated photons are shown for no dispersion engineering (a), matched group velocities (b) and in the case of additionally cancelled GVD (c). For each case, the accumulated phases during the PDC processes on the signal and idler photons are shown in the spectral domain together with the marginal spectral distribution. The joint spectral amplitude is depicted as the overlap of pump (blue) and phasematching (yellow) function, resulting in different outputs for the signal and idler spectra (red). A change in the temperature shifts the phase matching function as depicted.}
    \label{fig:concept}
\end{figure}

The consequences of GV matching and GVD cancellation on the generated signal and idler fields are depicted in Fig.\,\ref{fig:concept}. Here, we illustrate the resulting spectral shape and phase of the generated non-degenerate bi-photons and joint spectral amplitude from samples with optimized periodic poling and; no dispersion engineering (a), including GV matching (b), with simultaneous GV matching and GVD cancellation (c). In the case of no dispersion engineering, signal and idler experience a walk-off due to different group velocities at the generated wavelengths which leads to a different linear phase for the two colors. This walk-off can be prevented by matching the group velocities such that signal and idler emerge at the same time as illustrated in Fig.\,\ref{fig:concept} b). In that case, the photons obtain, to leading order, a quadratic phase and the marginal output spectra are broadened due to the larger overlap of phase matching and pump function. Finally, engineering the source to achieve both GV matching and GVD cancellation (Fig.\,\ref{fig:concept} c)) results in a bi-photon state whose phase profile is dominated by third order effects. In the resulting JSA a further broadened overlap of the phase matching and pump function is observed, and thus the generation of ultra-broadband bi-photons. 

\section{Experimental Setup}

The source is based on waveguides produced in z-cut LiNbO$_3$ fabricated in-house by in-diffusion of photo-lithographically patterned titanium strips with widths of $18\,\mathrm{\upmu m}$, $20\,\mathrm{\upmu m}$, and $22\,\mathrm{\upmu m}$. To allow for maximum flexibility, the sample features poling periods ranging from $5.8\,\mathrm{\upmu m}$ to $6.3\,\mathrm{\upmu m}$. We realize a type II phase-matching that allows to exploit the birefringence of the material in order to achieve simultaneous GV matching and GVD cancellation. The sample has a length of $40\,\mathrm{mm}$ and comprises 15 groups of three poled waveguides each. The mask layout is shown in Appendix 7.1.

We characterize the propagation losses of the samples by using a low-finesse Fabry-Pérot method \cite{Sohler1984}. We measured average losses of around $0.2\,\mathrm{dB/cm}$ for TM-polarized light at $3000\,\mathrm{nm}$, TE polarization has also been measured and experiences lower losses due to the waveguide geometry. A typical measurement for a whole waveguide chip is presented in Fig.\,\ref{fig:App_losses} in Appendix 7.2.

\begin{figure}[ht!]
    \centering
    \includegraphics[trim={0cm, 0cm, 0cm, 0cm},clip,width=\linewidth]{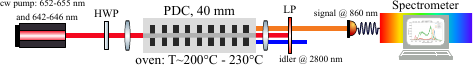}
    \caption{The setup to measure the generated signal spectra at a central wavelength of $860\,\mathrm{nm}$ from the nonlinear waveguide pumped by one of two tunable cw lasers. The polarization of the pump laser is set by a half wave plate (HWP) and the signal photons are filtered by a $735\,\mathrm{nm}$ long pass filter (LP) before detection to eliminate residual pump light.}
    \label{fig:setup}
\end{figure}

The setup to measure the generated signal spectrum is depicted in Fig.\,\ref{fig:setup}. We use one of two different external cavity diode lasers (TOPTICA DL pro) as pump lasers to address working points at different poling periods and waveguide temperatures across the sample. These lasers can be tuned from $652\,\mathrm{nm}$ to $655\,\mathrm{nm}$ and from $642\,\mathrm{nm}$ to $646\,\mathrm{nm}$, respectively. We monitor the wavelength of the lasers with the same spectrometer that we use for measuring the PDC spectra. For the first laser the waveguide sample is heated to about $230\,^{\circ}\mathrm{C}$ to achieve phase matching while a temperature of around $200\,^{\circ}\mathrm{C}$ is sufficient for the second laser. We designed the sample to operate at these temperatures to mitigate photo refractive effects \cite{Villarroel10}. We set the sample temperature with a home-built copper oven via a resistive heating cartridge which is driven by a temperature controller with integrated PID loop (Oxford Instruments MercuryiTC). The generated signal photons at a central wavelength of $860\,\mathrm{nm}$ are separated from the pump field by a $735\,\mathrm{nm}$ long-pass filter, coupled into a single mode fibre and detected using a single-photon sensitive spectrometer (Andor Shamrock SR-500i spectrograph with Newton 970P EMCCD-camera).

\section{Source Characterization}

\begin{figure}[ht!]
    \centering
    \includegraphics[trim={0cm, 0cm, 0cm, 0cm},clip,width=\linewidth]{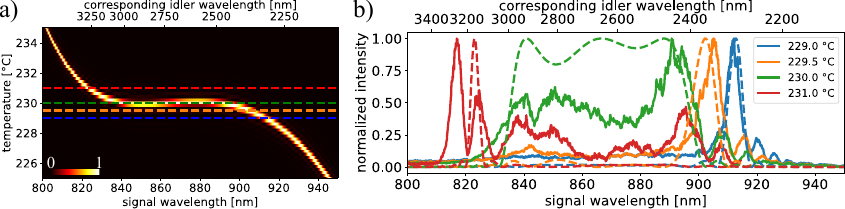}
    \caption{a) The simulated generated spectra at the design pump wavelength of $652.3\,\mathrm{nm}$ for different temperatures. The generated signal (idler) wavelengths are indicated on the bottom (top) x-axis. Colored, dashed lines show the temperature conditions at which spectra were measured experimentally. b) Measured signal spectra at different temperatures (solid, color lines) and simulated corresponding spectra (dashed, colored lines).}
    \label{fig:spectra}
\end{figure}

In order to achieve the desired broadband emission, the system needs to be precisely tuned to the working point, i.e. good dispersion engineering requires a reliable model that also captures higher order effects in the phase mismatch. We are therefore comparing our developed simulations against the measurements not only at the ideal working point but also at detuned pump wavelengths. To this end, we characterize our source in terms of changes in the spectral emission when varying the temperature of the waveguide and thus its refractive index at various pump wavelengths. The change in temperature primarily leads to a shift of the phase matching function with respect to the pump function in the JSA, c.f. Fig.\,\ref{fig:concept}. \\
We first start to investigate the temperature tuning with the pump laser set to the design wavelength of $652.3\,\mathrm{nm}$. In Fig.\,\ref{fig:spectra}, the simulated spectra at different temperatures around the working point are compared to the measured ones. The simulation in Fig.\,\ref{fig:spectra} a) shows the resulting spectrum of the signal photons for varying temperatures on the y-axis. The corresponding idler wavelengths are indicated on the top x-axis. It can be seen that the signal emission shifts from longer towards shorter wavelengths for increasing temperatures. Furthermore, a broadband emission is observed only for a specific temperature of $230\,^{\circ}\mathrm{C}$. At temperatures below and above this working point, the bandwidth of the emitted PDC decreases drastically. The measurements shown in Fig.\,\ref{fig:spectra} b) clearly confirm this behaviour. In this figure, the measured, normalized signal spectra are plotted together with the simulated spectra along the corresponding cuts on the temperature axis in the simulations, indicated via dashed lines of the corresponding colors. Both, the relative shift and the change in bandwidth are in good agreement. However, a systematic offset of around $40\,\mathrm{K}$ in the temperature and $2.4\,\mathrm{nm}$ in the pump wavelength had to be introduced in this and the following simulations. This deviation can be attributed to inaccuracies in our Sellmeier equations that are used to model the process. These Sellmeier equations are obtained by modelling the waveguide with the ideal fabrication parameters and are therefore also influenced by fabrication tolerances. However, these deviations only affect the first order terms, while the higher order terms are captured accurately.\\
At this optimal operation point the source covers wavelengths from $2400\,\mathrm{nm}$ to $3000\,\mathrm{nm}$, corresponding to a  spectral bandwidth of $25\,\mathrm{THz}$. Based on considerations from earlier work \cite{Roeder2024}, we extract a Fourier limited correlation time of less than $25\,\mathrm{fs}$ from this spectrum, which enables high resolution ultra-fast spectroscopic applications.

\begin{figure}[ht!]
    \centering
    \includegraphics[width=\linewidth]{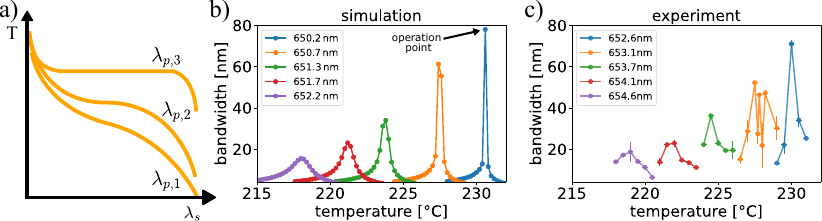}
    \caption{Change of the generated signal bandwidth for different waveguide temperatures at varying pump wavelengths above the design wavelength. a) A sketch illustrating the change in the temperature tuning for different pump wavelengths $\lambda_{p,1}>\lambda_{p,2}>\lambda_{p,3}$ that shows an increase in bandwidth closer to the design wavelength $\lambda_{p,3}$. b) Bandwidth of the signal spectrum at different temperatures and pump wavelengths calculated from simulations. c) The experimental spectra show the same increase in spectral bandwidth for decreasing pump wavelengths.}
    \label{fig:bandwidth_pump}
\end{figure}

In contrast to sources with only GV matching, for which a suitable operation point can be found for many pump wavelengths \cite{Pollmann2024}, in our case there is only one single operation pump wavelength for a given poling period due to the added constraint of GVD cancellation. Pump wavelengths that are offset from the design wavelength show a distinctive temperature tuning behaviour that has to be taken into account when operating such a source. This can be seen in Fig. \ref{fig:schematic_pm} in the Appendix where the pump wavelengths are set below, at, and above the design wavelength.\\
For a pump wavelength that is lower than the design wavelength, only a limited increase in emission bandwidth can be reached via temperature tuning. This behaviour is illustrated in Fig.\,\ref{fig:bandwidth_pump}. The simulation (Fig.\,\ref{fig:bandwidth_pump} b)) uses the temperature and pump wavelength offset that was identified in Fig.\,\ref{fig:spectra}. From these simulations, the bandwidth of the signal spectrum was extracted using a full width at $80\,\%$ of the maximum. This criterion has been chosen to avoid the influence of side lobes in the non-perfect experimental phase matching. Indeed, two data points in Fig.\,\ref{fig:bandwidth_pump} c) show a more narrow bandwidth than expected for a pump wavelength of $653\,\mathrm{nm}$ which is caused by phase matching side lobes being evaluated instead of the main peak. The error bars consider an uncertainty in the set temperature of $0.2\,\mathrm{K}$ and a $5\,\%$ interval of the signal height for the width estimation.\\
It can be seen that the temperature required to achieve the maximum bandwidth is lower for higher pump wavelengths. As the pump wavelength is reduced and ultimately matches with the design wavelength, the maximal bandwidth increases while the temperature range over which the increase in bandwidth happens decreases. The same behaviour can be observed for the experimental results. An implication of these measurements is that the accepted temperature range can be extended at the cost of maximal bandwidth for a detuned pump wavelength, which might be useful for applications where a reduced bandwidth is suitable with less critical temperature stability.

\begin{figure}[ht!]
    \centering
    \includegraphics[width=\linewidth]{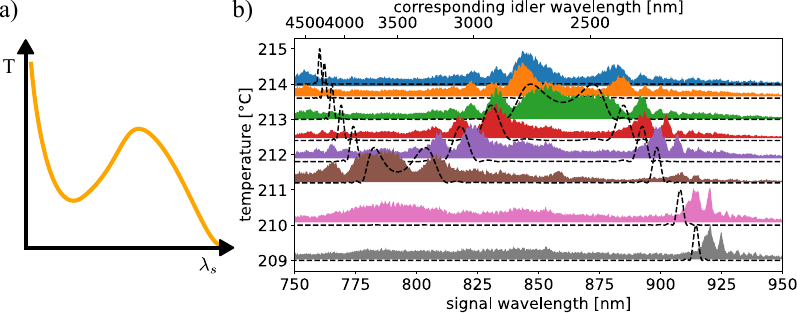}
    \caption{a) Schematic representation of the temperature tuning curve for a pump wavelength below the design wavelength. b) The change in the signal spectrum with temperature for a pump wavelength of $644.25\,\mathrm{nm}$. Each spectrum is normalized to its maximum. This behaviour matches with the expected shape of the temperature tuning curve for a too low pump wavelength as shown in the Appendix. The dotted curves represent the simulations at the corresponding temperatures.}
    \label{fig:pump_waterfall}
\end{figure}

Finally, we investigate the behaviour of our source at a pump wavelength that is lower than the design wavelength. As depicted in Fig.\,\ref{fig:pump_waterfall} a), increasing the waveguide temperature first leads to a broadband emission at low wavelength, followed by one at longer wavelength at higher temperatures. This peculiar behaviour can be reproduced by the simulations that also show two regions of more broadband emission and are depicted by the dotted lines in the figure. Although there are differences in the structure between the measured results and the simulations, the two qualitatively match in their general behaviour. Deviations in the waveguide temperature are the most likely cause for this. The temperatures in Fig.\,\ref{fig:pump_waterfall} b) are lower than the previous ones as the second pump laser with a smaller wavelength of $644.25\,\mathrm{nm}$ has been used to generate these signal spectra. The full picture of the temperature tuning behaviours for different pump wavelengths is presented in Fig.\,\ref{fig:schematic_pm} in the Appendix.\\

Thus, characterization of the temperature tuning behaviour of our source for different pump wavelengths shows that careful source design is required to reach the desired broadband emission. The observed behaviour has been captured by our bespoke model and agrees very well with the presented experimental data. A high precision in modelling and experimentally reaching the working point is needed due to the fact that for a given poling period, the correct pump wavelength and waveguide temperature have to be chosen in order to fulfill both GV matching and the cancellation of GVD for the signal and idler fields. This understanding is crucial for future experiments that involve this or similar sources.

Despite expecting a high brightness comparable to the one reported in \cite{Pollmann2024}, the actual brightness of the source could not be measured due to the lack of single-photon detectors in the mid-infrared which prohibits coincidence detection, as typical detectors show a high thermal noise level and are not sensitive to single photons. However, a lower bound of $5\cdot10^3 \frac{\mathrm{counts}}{\mathrm{s} \cdot \mathrm{mW} \cdot \mathrm{GHz}}$ can be estimated by measurements of the signal rate. This value makes the source at least comparable to or even brighter than bulk PDC sources. For more details, see Appendix 7.4.\\

\section{Discussion and Outlook}

We have presented the design and characterization of a broadband PDC source that generates bi-photons at non-degenerate wavelengths, one in the NIR at a central wavelength of $860\,\mathrm{nm}$ and the other in the MIR at $2800\,\mathrm{nm}$. For this purpose, we fabricated a $40\,\mathrm{mm}$ long waveguide chip with waveguides of widths $18\,\mathrm{\upmu m}$, $20\,\mathrm{\upmu m}$, and $22\,\mathrm{\upmu m}$ and verified guiding of light at $3000\,\mathrm{nm}$ with low losses ($<0.2\mathrm{dB}/\mathrm{cm}$). The spectral bandwidth of the generated bi-photons reaches more than $25\,\mathrm{THz}$ for both signal and idler which is achieved via simultaneous GV matching and GVD cancellation. The theoretically predicted dependence of the output spectrum on the temperature and pump wavelength was verified. Furthermore, we estimated the achieved brightness to exceed $5\cdot10^3 \frac{\mathrm{counts}}{\mathrm{s} \cdot \mathrm{mW} \cdot \mathrm{GHz}}$ , which is at least comparable to similar PDC sources.\\
Our source offers great potential for applications in ultrafast quantum spectroscopy and sensing. Due to the broad spectral coverage as well as the bright emission from the confinement over $40\,\mathrm{mm}$ length in a waveguide, this source overcomes the typical trade-off between brightness and bandwidth. Furthermore, non-degenerate emission allows for utilizing the source in the context of nonlinear interferometers to perform measurements with undetected photons where an object under test can be probed by the photons in the mid IR while detection happens in the near IR.

\section{Acknowledgements}

F.R. is part of the Max Planck School of Photonics supported by the German Federal Ministry of Education and Research (BMBF), the Max Planck Society, and the Fraunhofer Society. We acknowledge financial support from the Federal Ministry of Education and Research (BMBF) via the grant agreement no. 13N16352 (E2TPA). This project has received funding from the European Union’s Horizon Europe research and innovation programme under grant agreement no. 101070700 (MIRAQLS).

\section{Appendix}

\subsection{Sample Layout}

\begin{figure}[ht!]
    \centering
    \includegraphics[trim={0cm, 0cm, 0cm, 0cm},clip,width=\linewidth]{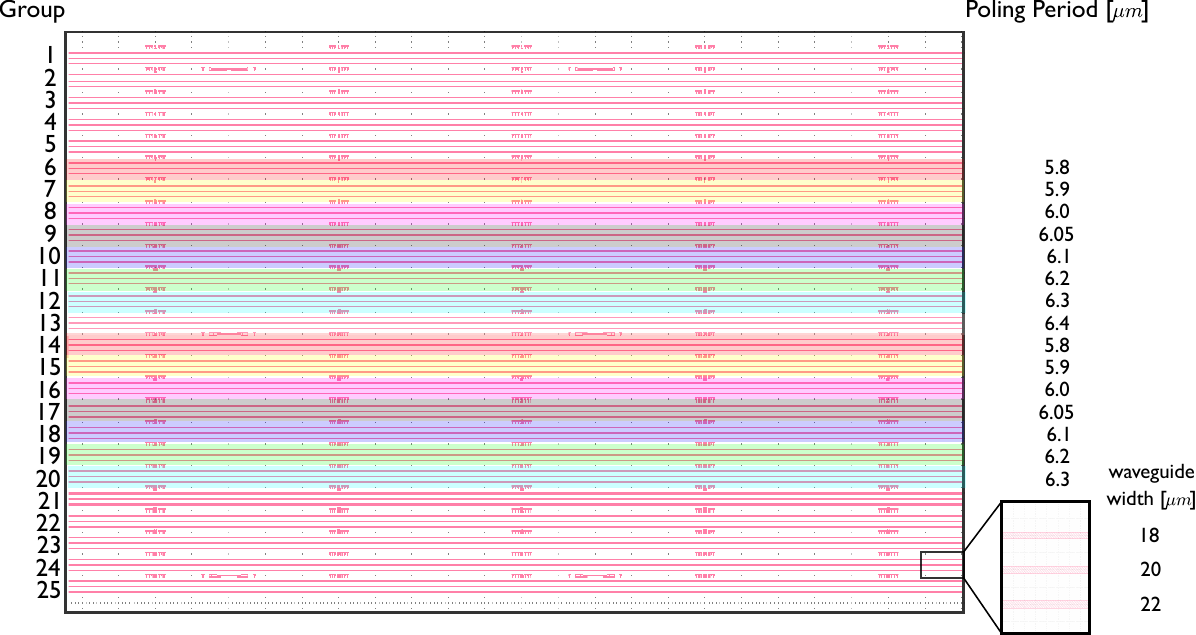}
    \caption{The layout of the waveguide photomask with marked poling periods that are assigned to the corresponding groups of waveguides. Each of these groups contains waveguides of widths 18, 20 and 22\,$\mathrm{\upmu m}$.}
    \label{fig:App_mask}
\end{figure}

The photo-mask containing the waveguides and poling structure is presented in Fig.\,\ref{fig:App_mask}

\subsection{Linear optical losses}

\begin{figure}[ht!]
    \centering
    \includegraphics[trim={0cm, 0cm, 0cm, 0cm},clip,width=0.8\linewidth]{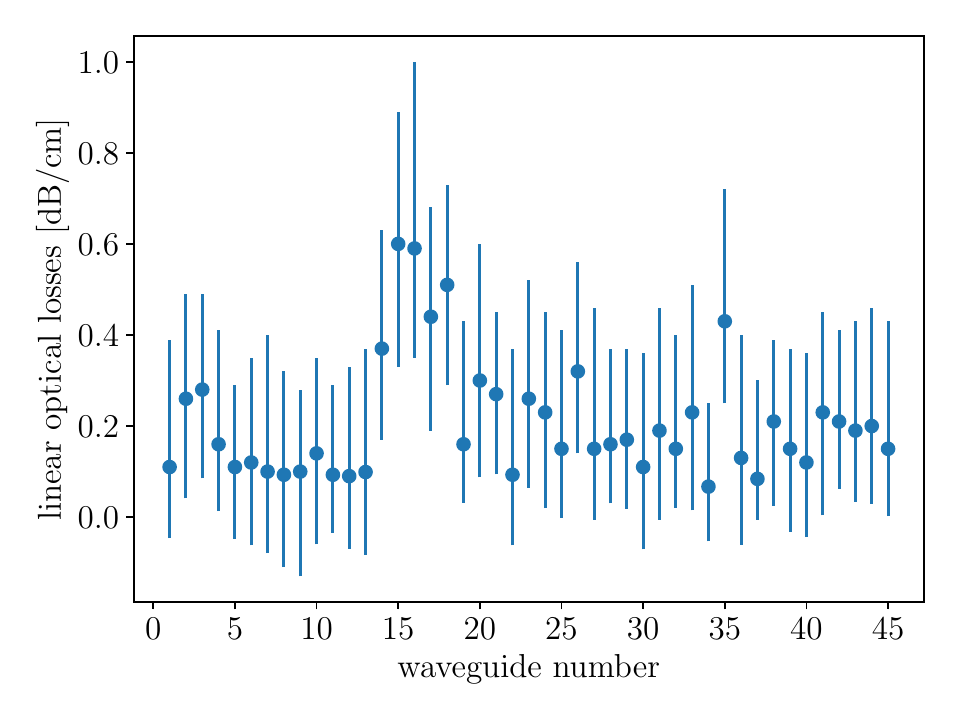}
    \caption{The linear optical losses at $3000\,\mathrm{nm}$ and TM polarization are presented for different waveguides on one of the fabricated waveguide chips.}
    \label{fig:App_losses}
\end{figure}

The measured linear optical losses with the Fabry-Pérot method, c.f. \cite{Sohler1984}, at $3000\,\mathrm{nm}$ in TM polarization for the waveguide chip used above with 45 waveguides are shown in Fig.\,\ref{fig:App_losses}. The average losses are below $0.2\,\mathrm{dB/cm}$. From these measurements, the best waveguides are chosen for further experiments. The error bars are associated with the thermal noise of the used mid IR detector.

\subsection{Temperature characteristics}

\begin{figure}[ht!]
    \centering
    \includegraphics[trim={0cm, 0cm, 0cm, 0cm},clip,width=\linewidth]{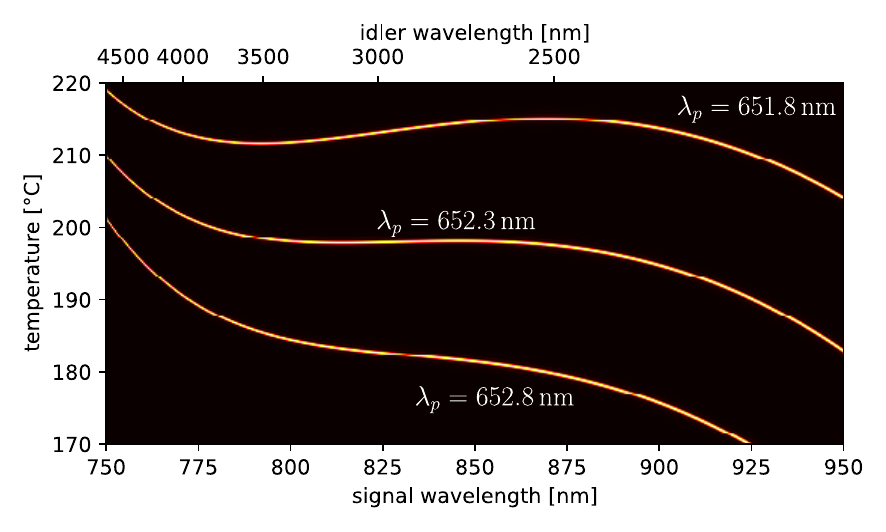}
    \caption{The resulting signal spectrum with varying temperature when the pump wavelength is set above ($652.8\,\mathrm{nm}$), at ($652.3\,\mathrm{nm}$) or below ($651.8\,\mathrm{nm}$) the design wavelength.}
    \label{fig:schematic_pm}
\end{figure}

An overview of the three possible temperature tuning behaviours for varying pump wavelengths around the design wavelength is shown in Fig.\,\ref{fig:schematic_pm}. When the pump wavelength is larger than the optimal value, no or only a less broadband emission is reached as no region parallel to the x-axis forms during temperature tuning. For the design wavelength, this region is present and spans more than $50\,\mathrm{nm}$ in the given example plot. If the pump wavelength is chosen too low, the formation of up to 3 peaks in the signal spectrum is possible for a specific temperature. Furthermore, two regions of less broadband emission form while increasing the temperature, first at shorter wavelengths around $800\,\mathrm{nm}$ and later at longer wavelengths around $875\,\mathrm{nm}$.

\subsection{Brightness estimation}

A lower bound for the brightness of the source is estimated from the counts measured when detecting the signal field using an avalanche photodiode (APD). Despite not being able to measure coincidences between signal and idler photons, which would allow to differentiate counts caused from the bi-photons from background and noise, the simultaneous measurement of photon counts and the spectra make it possible to calculate the bi-photon rate, normalized to the pump power and spectral width. This estimation does not contain coupling efficiencies to the detection and therefore only provides a lower bound. We calibrate the counts on the spectrograph against the ones on the APD by measuring the same signal with sufficient attenuation on the APD. This allows one to substract background and fluorescence counts from the spectrum and calculate a lower bound for the number of generated pairs. The brightness measured in this way, corrected for fluorescence and background noise, is $5 \cdot 10^{3}\,\frac{\mathrm{counts}}{\mathrm{s} \cdot \mathrm{mW} \cdot \mathrm{GHz}}$. If one takes into account the estimated coupling efficiency of $20\,\%$, this leads to a number that is brighter than most bulk sources and comparable to other waveguide sources, c.f. \cite{Pollmann2024}, while achieving a large bandwidth.\\
\newpage

\printbibliography

@PREAMBLE{
 "\providecommand{\noopsort}[1]{}" 
 # "\providecommand{\singleletter}[1]{#1}%" 
}

@misc{Williams2024,
      title={Ultra-Short Pulse Biphoton Source in Lithium Niobate Nanophotonics at 2 \textmu m}, 
      author={James Williams and Rajveer Nehra and Elina Sendonaris and Luis Ledezma and Robert M. Gray and Ryoto Sekine and Alireza Marandi},
      year={2024},
      eprint={2402.05163},
      archivePrefix={arXiv},
      primaryClass={physics.optics}
}

@article{Kaufmann2022,
author = {Paul Kaufmann and Helen M Chrzanowski and Aron Vanselow and Sven Ramelow},
journal = {Opt. Express},
number = {4},
pages = {5926--5936},
publisher = {Optica Publishing Group},
title = {Mid-IR spectroscopy with NIR grating spectrometers},
volume = {30},
month = {Feb},
year = {2022},
url = {https://opg.optica.org/oe/abstract.cfm?URI=oe-30-4-5926},
doi = {10.1364/OE.442411},
}

@article{Paterova2018,
doi = {10.1088/1367-2630/aab5ce},
url = {https://dx.doi.org/10.1088/1367-2630/aab5ce},
year = {2018},
month = {apr},
publisher = {IOP Publishing},
volume = {20},
number = {4},
pages = {043015},
author = {Anna Paterova and Hongzhi Yang and Chengwu An and Dmitry Kalashnikov and Leonid Krivitsky},
title = {Measurement of infrared optical constants with visible photons},
journal = {New Journal of Physics},
}

@article{Lindner2021,
author = {Chiara Lindner and Jachin Kunz and Simon J. Herr and Sebastian Wolf and Jens Kie{\ss}ling and Frank K\"{u}hnemann},
journal = {Opt. Express},
number = {3},
pages = {4035--4047},
publisher = {Optica Publishing Group},
title = {Nonlinear interferometer for Fourier-transform mid-infrared gas spectroscopy using near-infrared detection},
volume = {29},
month = {Feb},
year = {2021},
url = {https://opg.optica.org/oe/abstract.cfm?URI=oe-29-3-4035},
doi = {10.1364/OE.415365},
}

@ARTICLE{Kviatkovsky2020,
author = {Inna Kviatkovsky  and Helen M. Chrzanowski  and Ellen G. Avery  and Hendrik Bartolomaeus  and Sven Ramelow },
title = {Microscopy with undetected photons in the mid-infrared},
journal = {Science Advances},
volume = {6},
number = {42},
pages = {eabd0264},
year = {2020},
doi = {10.1126/sciadv.abd0264},
}

@article{Chen2017,
author = {Changjia Chen and Eric Y. Zhu and Arash Riazi and Alexey V. Gladyshev and Costantino Corbari and Morten Ibsen and Peter G. Kazansky and Li Qian},
journal = {Opt. Express},
number = {19},
pages = {22667--22678},
publisher = {Optica Publishing Group},
title = {Compensation-free broadband entangled photon pair sources},
volume = {25},
month = {Sep},
year = {2017},
url = {https://opg.optica.org/oe/abstract.cfm?URI=oe-25-19-22667},
doi = {10.1364/OE.25.022667},
}

@ARTICLE{Lemos2014,
   author       = "Lemos, Gabriela Barreto and Borish, Victoria and Cole, Garrett D. and Ramelow, Sven and Lapkiewicz, Radek and Zeilinger, Anton",
   year         = "2014",
   journal      = "Nature",
   volume       = "512",
   pages        = "409-412",
   title        = "Quantum imaging with undetected photons",
}

@ARTICLE{Toepfer2022,
author = {Sebastian Töpfer  and Marta Gilaberte Basset  and Jorge Fuenzalida  and Fabian Steinlechner  and Juan P. Torres  and Markus Gräfe },
title = {Quantum holography with undetected light},
journal = {Sci. Adv.},
volume = {8},
number = {2},
pages = {eabl4301},
year = {2022},
doi = {10.1126/sciadv.abl4301},
}

@article{Riazi2019,
author = {Arash Riazi and Eric Y. Zhu and Changjia Chen and Alexey V. Gladyshev and Peter G. Kazansky and Li Qian},
journal = {Opt. Lett.},
number = {6},
pages = {1484--1487},
publisher = {Optica Publishing Group},
title = {Alignment-free dispersion measurement with interfering biphotons},
volume = {44},
month = {Mar},
year = {2019},
url = {https://opg.optica.org/ol/abstract.cfm?URI=ol-44-6-1484},
doi = {10.1364/OL.44.001484},
}

@article{Mukamel2020,
doi = {10.1088/1361-6455/ab69a8},
url = {https://dx.doi.org/10.1088/1361-6455/ab69a8},
year = {2020},
month = {mar},
publisher = {IOP Publishing},
volume = {53},
number = {7},
pages = {072002},
author = {Shaul Mukamel and Matthias Freyberger and Wolfgang Schleich and Marco Bellini and Alessandro Zavatta and Gerd Leuchs and Christine Silberhorn and Robert W Boyd and Luis Lorenzo Sánchez-Soto and André Stefanov and Marco Barbieri and Anna Paterova and Leonid Krivitsky and Sharon Shwartz and Kenji Tamasaku and Konstantin Dorfman and Frank Schlawin and Vahid Sandoghdar and Michael Raymer and Andrew Marcus and Oleg Varnavski and Theodore Goodson and Zhi-Yuan Zhou and Bao-Sen Shi and Shahaf Asban and Marlan Scully and Girish Agarwal and Tao Peng and Alexei V Sokolov and Zhe-Dong Zhang and M Suhail Zubairy and Ivan A Vartanyants and Elena del Valle and Fabrice Laussy},
title = {Roadmap on quantum light spectroscopy},
journal = {Journal of Physics B: Atomic, Molecular and Optical Physics}
}

@article{Roeder2024,
  title = {Measurement of Ultrashort Biphoton Correlation Times with an Integrated Two-Color Broadband $\mathrm{SU}(1,1)$-Interferometer},
  author = {Roeder, F. and Pollmann, R. and Stefszky, M. and Santandrea, M. and Luo, K.-H. and Quiring, V. and Ricken, R. and Eigner, C. and Brecht, B. and Silberhorn, C.},
  journal = {PRX Quantum},
  volume = {5},
  pages = {020350},
  numpages = {11},
  year = {2024},
  month = {May},
  publisher = {American Physical Society},
  doi = {10.1103/PRXQuantum.5.020350},
  url = {https://link.aps.org/doi/10.1103/PRXQuantum.5.020350}
}

@article{Chekhova2016,
author = {M. V. Chekhova and Z. Y. Ou},
journal = {Adv. Opt. Photon.},
number = {1},
pages = {104--155},
publisher = {Optica Publishing Group},
title = {Nonlinear interferometers in quantum optics},
volume = {8},
month = {Mar},
year = {2016},
url = {https://opg.optica.org/aop/abstract.cfm?URI=aop-8-1-104},
doi = {10.1364/AOP.8.000104},
}

@article{Manceau2017,
  title = {Detection Loss Tolerant Supersensitive Phase Measurement with an SU(1,1) Interferometer},
  author = {Manceau, Mathieu and Leuchs, Gerd and Khalili, Farid and Chekhova, Maria},
  journal = {Phys. Rev. Lett.},
  volume = {119},
  pages = {223604},
  numpages = {5},
  year = {2017},
  month = {Nov},
  publisher = {American Physical Society},
  doi = {10.1103/PhysRevLett.119.223604},
  url = {https://link.aps.org/doi/10.1103/PhysRevLett.119.223604}
}

@article{Pollmann2024,
author = {Ren\'{e} Pollmann and Franz Roeder and Victor Quiring and Raimund Ricken and Christof Eigner and Benjamin Brecht and Christine Silberhorn},
journal = {Opt. Express},
number = {14},
pages = {23945--23955},
publisher = {Optica Publishing Group},
title = {Integrated, bright broadband, two-colour parametric down-conversion source},
volume = {32},
month = {Jul},
year = {2024},
url = {https://opg.optica.org/oe/abstract.cfm?URI=oe-32-14-23945},
doi = {10.1364/OE.522549},
}

@ARTICLE{Sohler1984,
  title = {Loss in low-finesse Ti:LiNbO3 optical waveguide resonates},
  author = {Sohler, W and Regener, R.},
  journal = {Appl. Phys. B},
  pages = {144-146},
  year = {1984}
}

@ARTICLE{Tanzilli2001,
   author = {S. Tanzilli},
   author = {H. De Riedmatten},
   author = {W. Tittel},
   author = {H. Zbinden},
   author = {P. Baldi},
   author = {M. De Micheli},
   author = {D.B. Ostrowsky},
   author = {N. Gisin},
   ISSN = {0013-5194},
   language = {English},
   title = {Highly efficient photon-pair source using periodically poled lithium niobate waveguide},
   journal = {Electronics Letters},   
   volume = {37},
   year = {2001},
   month = {January},
   pages = {26-28(2)},
   publisher ={Institution of Engineering and Technology},
   copyright = {© IEE},
   url = {https://digital-library.theiet.org/content/journals/10.1049/el_20010009}
}

@misc{katamadze2024,
      title={Broadband biphoton source for quantum optical coherence tomography based on a Michelson interferometer}, 
      author={Konstantin Katamadze and Anna Romanova and Denis Chupakhin and Alexander Pashchenko and Sergei Kulik},
      year={2024},
      eprint={2401.17836},
      archivePrefix={arXiv},
      primaryClass={quant-ph},
      url={https://arxiv.org/abs/2401.17836}, 
}

@article{Ndagano2022,
author = {B. Ndagano, H. Defienne, D. Branford et al.},
journal = {Nat. Photon.},
number = {16},
pages = {384--389},
title = {Quantum microscopy based on Hong-Ou-Mandel interference},
year = {2022},
}

@article{Yepiz-Graciano2020,
author = {Pablo Yepiz-Graciano and Al\'{i} Michel Angulo Mart\'{i}nez and Dorilian Lopez-Mago and Hector Cruz-Ramirez and Alfred B. U'Ren},
journal = {Photon. Res.},
number = {6},
pages = {1023--1034},
publisher = {Optica Publishing Group},
title = {Spectrally resolved Hong-Ou-Mandel interferometry for quantum-optical coherence tomography},
volume = {8},
month = {Jun},
year = {2020},
url = {https://opg.optica.org/prj/abstract.cfm?URI=prj-8-6-1023},
}

@ARTICLE{Uren2005,
   author       = "U'Ren, A. B. and Silberhorn, C. and Banaszek, K. and Walmsley, I. and Erdmann, R. and Grice, W. and Raymer, M. G.", 
   title        = "Generation of Pure-State Single-Photon Wavepackets by Conditional Preparation Based on Spontaneous Parametric Downconversion", 
   journal      = "Las. Phys.", 
   volume       = "15", 
   pages        = "146 - 161", 
   year         = "2005", 
}

@article{Kolenderska2020,
author = {Sylwia M. Kolenderska and Fr\'{e}d\'{e}rique Vanholsbeeck and Piotr Kolenderski},
journal = {Opt. Express},
number = {20},
pages = {29576--29589},
publisher = {Optica Publishing Group},
title = {Fourier domain quantum optical coherence tomography},
volume = {28},
month = {Sep},
year = {2020},
url = {https://opg.optica.org/oe/abstract.cfm?URI=oe-28-20-29576},
doi = {10.1364/OE.399913},
}

@article{Smirnov2024,
author = {Maksim A. Smirnov and Ilya V. Fedotov and Anastasia M. Smirnova and Albert F. Khairullin and Andrei B. Fedotov and Sergey A. Moiseev},
journal = {Opt. Lett.},
number = {14},
pages = {3838--3841},
publisher = {Optica Publishing Group},
title = {Bright ultra-broadband fiber-based biphoton source},
volume = {49},
month = {Jul},
year = {2024},
url = {https://opg.optica.org/ol/abstract.cfm?URI=ol-49-14-3838},
doi = {10.1364/OL.524201},
}

@article{Tashima_24,
author = {Toshiyuki Tashima and Yu Mukai and Masaya Arahata and Norihide Oda and Mamoru Hisamitsu and Katsuhiko Tokuda and Ryo Okamoto and Shigeki Takeuchi},
journal = {Optica},
number = {1},
pages = {81--87},
publisher = {Optica Publishing Group},
title = {Ultra-broadband quantum infrared spectroscopy},
volume = {11},
month = {Jan},
year = {2024},
url = {https://opg.optica.org/optica/abstract.cfm?URI=optica-11-1-81},
doi = {10.1364/OPTICA.504450},
}

@ARTICLE{Ramelow2019,
   author       = "A. Vanselow et al.", 
   title        = "Ultra-broadband {SPDC} for spectrally far separated photon pairs", 
   journal      = "Opt. Lett.", 
   volume       = "44", 
   pages        = "4638-4641", 
   year         = "2019", 
}

@article{Dayan2004,
  title = {Two Photon Absorption and Coherent Control with Broadband Down-Converted Light},
  author = {Dayan, Barak and Pe'er, Avi and Friesem, Asher A. and Silberberg, Yaron},
  journal = {Phys. Rev. Lett.},
  volume = {93},
  pages = {023005},
  numpages = {4},
  year = {2004},
  month = {Jul},
  publisher = {American Physical Society},
  doi = {10.1103/PhysRevLett.93.023005},
  url = {https://link.aps.org/doi/10.1103/PhysRevLett.93.023005}
}

@article{Shaked2014,
  title = {Observing the nonclassical nature of ultra-broadband bi-photons at ultrafast speed},
  author = {Y. Shaked and R. Pomerantz and R. Z. Vered and A. Pe'er},
  journal = {New J. Phys.},
  volume = {16},
  pages = {053012},
  year = {2014},
}

@article{Raymer2022,
  title = {Theory of two-photon absorption with broadband squeezed vacuum},
  author = {Raymer, Michael G. and Landes, Tiemo},
  journal = {Phys. Rev. A},
  volume = {106},
  pages = {013717},
  numpages = {16},
  year = {2022},
  month = {Jul},
  publisher = {American Physical Society},
  doi = {10.1103/PhysRevA.106.013717},
  url = {https://link.aps.org/doi/10.1103/PhysRevA.106.013717}
}

@article{Zhu2013,
author = {E. Y. Zhu and Z. Tang and L. Qian and L. G. Helt and M. Liscidini and J. E. Sipe and C. Corbari and A. Canagasabey and M. Ibsen and P. G. Kazansky},
journal = {Opt. Lett.},
number = {21},
pages = {4397--4400},
publisher = {Optica Publishing Group},
title = {Poled-fiber source of broadband polarization-entangled photon pairs},
volume = {38},
month = {Nov},
year = {2013},
url = {https://opg.optica.org/ol/abstract.cfm?URI=ol-38-21-4397},
doi = {10.1364/OL.38.004397},
}

@article{Smith2009,
author = {Brian J. Smith and P. Mahou and Offir Cohen and J. S. Lundeen and I. A. Walmsley},
journal = {Opt. Express},
number = {26},
pages = {23589--23602},
publisher = {Optica Publishing Group},
title = {Photon pair generation in birefringent optical fibers},
volume = {17},
month = {Dec},
year = {2009},
url = {https://opg.optica.org/oe/abstract.cfm?URI=oe-17-26-23589},
doi = {10.1364/OE.17.023589},
}

@ARTICLE{Grice1997,
  title = {Spectral information and distinguishability in type-II down-conversion with a broadband pump},
  author = {Grice, W. P. and Walmsley, I. A.},
  journal = {Phys. Rev. A},
  volume = {56},
  pages = {1627--1634},
  numpages = {0},
  year = {1997},
  month = {Aug},
  publisher = {American Physical Society},
  doi = {10.1103/PhysRevA.56.1627},
  url = {https://link.aps.org/doi/10.1103/PhysRevA.56.1627}
}

@article{Villarroel10,
author = {J. Villarroel and J. Carnicero and F. Luedtke and M. Carrascosa and A. Garc\'{i}a-Caba\~{n}es and J. M. Cabrera and A. Alcazar and B. Ramiro},
journal = {Opt. Express},
number = {20},
pages = {20852--20861},
publisher = {Optica Publishing Group},
title = {Analysis of photorefractive optical damage in lithium niobate: application to planar waveguides},
volume = {18},
month = {Sep},
year = {2010},
url = {https://opg.optica.org/oe/abstract.cfm?URI=oe-18-20-20852},
doi = {10.1364/OE.18.020852},
}

@article{Fang24,
author = {Xiao-Xu Fang and Leiran Wang and He Lu},
journal = {Opt. Express},
number = {13},
pages = {22945--22954},
publisher = {Optica Publishing Group},
title = {Efficient generation of broadband photon pairs in shallow-etched lithium niobate nanowaveguides},
volume = {32},
month = {Jun},
year = {2024},
url = {https://opg.optica.org/oe/abstract.cfm?URI=oe-32-13-22945},
doi = {10.1364/OE.519265},
}

@article{Fraine2012,
author = {A. Fraine and O. Minaeva and D. S. Simon and R. Egorov and A. V. Sergienko},
journal = {Opt. Lett.},
number = {11},
pages = {1910--1912},
publisher = {Optica Publishing Group},
title = {Broadband source of polarization entangled photons},
volume = {37},
month = {Jun},
year = {2012},
url = {https://opg.optica.org/ol/abstract.cfm?URI=ol-37-11-1910},
doi = {10.1364/OL.37.001910},
}

@article{Landes2021,
  title = {Experimental feasibility of molecular two-photon absorption with isolated time-frequency-entangled photon pairs},
  author = {Landes, Tiemo and Allgaier, Markus and Merkouche, Sofiane and Smith, Brian J. and Marcus, Andrew H. and Raymer, Michael G.},
  journal = {Phys. Rev. Res.},
  volume = {3},
  pages = {033154},
  numpages = {9},
  year = {2021},
  month = {Aug},
  publisher = {American Physical Society},
  doi = {10.1103/PhysRevResearch.3.033154},
  url = {https://link.aps.org/doi/10.1103/PhysRevResearch.3.033154}
}

@article{Chekhova2019,
  title = {Microscale Generation of Entangled Photons without Momentum Conservation},
  author = {Okoth, C. and Cavanna, A. and Santiago-Cruz, T. and Chekhova, M. V.},
  journal = {Phys. Rev. Lett.},
  volume = {123},
  pages = {263602},
  numpages = {6},
  year = {2019},
  month = {Dec},
  publisher = {American Physical Society},
  doi = {10.1103/PhysRevLett.123.263602},
  url = {https://link.aps.org/doi/10.1103/PhysRevLett.123.263602}
}

@article{Yurke1986,
  title = {SU(2) and SU(1,1) interferometers},
  author = {Yurke, Bernard and McCall, Samuel L. and Klauder, John R.},
  journal = {Phys. Rev. A},
  volume = {33},
  pages = {4033--4054},
  numpages = {0},
  year = {1986},
  publisher = {American Physical Society},
  doi = {10.1103/PhysRevA.33.4033},
  url = {https://link.aps.org/doi/10.1103/PhysRevA.33.4033}
}

\end{document}